\documentclass[amsmath,amssymb,onecolumn]{revtex4}
\usepackage{amssymb}
\usepackage{amsfonts}
\usepackage{amsmath}
\usepackage{epstopdf}
\usepackage{color}
\usepackage{graphicx}
\usepackage{pdfpages}
\usepackage{hyperref}
\usepackage{cleveref}

\begin{document}
\title{Some global, analytical and topological properties of regular black holes}

\author{Pedro \surname{Bargueño}}
\email{pedro.bargueno@ua.es}
\affiliation{Departamento de F\'{i}sica Aplicada, Universidad de Alicante, Campus de San Vicente del Raspeig, E-03690 Alicante, Spain}

\date{\today}

\begin{abstract}
In this work we study regular black holes from a global perspective looking for evading some of the well-known singularity theorems by using their ``reverses". Then, model geometries for the slices of typical spherically symmetric, (locally) static four dimensional regular black hole solutions are described from both an analytical and a topological point of view. While the finiteness of both the scalar and Kretchmann curvature of the slices around the regular center determines the geometry of the core, the positive answer to the Poincaré conjecture assures that, under two assumptions, its topology is that of a three-sphere. However, in general, the cores are shown to be $S^3$, $H^3$, $\mathbb{R}\times S^1$ or $S^1\times S^2$, depending whether a de Sitter, anti de Sitter, Nariai or Bertotti-Robinson geometries are employed to describe the slices at the regular center. Then, a description of the aforementioned slices in terms of Seifert fibre spaces is given in order to show that the Euler characteristic of the bundle can be used to track the transition between the core of the regular black hole and the rest of the slices in most of the cases considered in the literature. After Geroch and Tipler's theorems are employed to study the consequences of topology change on regular black hole spacetimes, we show that Borde and Vilenkin's singularity theorem is used to restrict their possible types. We end by noting that Nariai cores can be safely used to construct regular black holes without topology changes.
\end{abstract}

\maketitle

\section{Introduction and purpose of the work}

Although a fully working theory of quantum gravity is not at hand, regular black holes can be used as phenomenological toy models in order to investigate possible ways of singularity resolutions. 
The seminal idea of Sakharov\cite{Sakharov1966TheMatterb} and Gliner\cite{Gliner1966AlgebraicMatterb}, proposing that singularities 
could be substituted by an inflationary equation of state 
({\it i. e.}, a de Sitter core), was firstly realized by Bardeen in 1968\cite{Bardeen1968}. Since then, although there have been a great development within the field, most of regular black hole solutions (without layers) were constructed following Bardeen's proposal. 
The first exact solution for a regular black hole was found by Ayon-Beato and García in 1998 using non-linear electrodynamics as the source, giving a big impulse to the field \cite{Ayon-Beato2000TheMonopole}.
In particular, Dymnikova\cite{Dymnikova1992VacuumHole,Dymnikova2002TheMass,Dymnikova2003SphericallyCenter,Dymnikova2004RegularRelativity} found an  asymptotically Schwarzschild regular black hole solution whose interior corresponds to an anisotropic fluid obeying a de Sitter equation of state.
Other regular black hole solutions with a de Sitter core were found by similar techniques \cite{Ayon-Beato2000TheMonopole,Lemos2011RegularCore,Balart2014RegularSource,Morales-Duran2016SimpleCorrection}, including regular black holes in different extended theories of gravity\cite{Aftergood2014MatterGravity,Rodrigues2016RegularElectrodynamics,Contreras2018ASolution}. Very recently, regular black holes with a Minkowski core have been recently reported \cite{Simpson2019RegularCores}, allowing thus to explore new possibilities such as the construction of thin-shell traversable wormholes \cite{Berry2020}. 
There are also regular black hole models which have been considered as dark matter candidates (see 
\cite{Rovelli2018,Dymni2020} for very recent reviews on the subject), including extremal configurations \cite{Freitas2018}.
In particular, regular black hole remnants, G-lumps and graviatoms can be considered heavy dark matter candidates with dark energy interiors
\cite{D2020} which can induce observational consequences, such as proton decay \cite{Dymni2015}. Interestingly, a very recent example of a classical mechanism giving place to a regular black hole without a de Sitter but 
a Nariai center has been introduced in \cite{Mariam} by the use of three-form fields. Within the quantum realm, similar regular black holes can also be formulated in loop quantum gravity \cite{LQC1,LQC2,LQC3,LQC4,LQC5} and, within a de Sitter core, in string theory-inspired corrections \cite{Nicolini2019}. Interestingly, regular stringy black holes without a de Sitter core have been recently reported \cite{Cano2019}. Finally, we would like draw attention on some intriguing results regarding the classical double-copy \cite{Monteiro2014} of regular black holes \cite{Easson2020}.

Although during the last years the number of works on regular black holes has been constantly increasing, much less has been said from the point of view of global techniques applied to these spacetimes. The first work explaining how to avoid Penrose's singularity theorem \cite{Penrose1965} to construct regular black holes is Borde's 1997 theorem
\cite{Borde1997RegularChange}. From this moment, the majority (if not all) the works on regular black holes referred to this theorem in order to justify the omnipresent de Sitter core. This is also the starting point of the so-called topology change in regular black holes, since the asymptotically flat region is usually assumed to be $\mathbb{R}\times S^2$ and the core is represented by $S^3$. Very recently, Melgarejo, Contreras and the author of the present work have shown \cite{Melgarejo2020} that topologies other than $S^3$ are admissible in spherically symmetric and static black holes with a Carter-Penrose diagram {\it à la} Reissner-Nordstr\"om, which essentially describes most of regular black hole models reported. In addition, Carballo-Rubio, Di Filippo, Liberati and Visser have classified all possible regular spherically symmetric geometries that may be realized in theories beyond general relativity as the result of
singularity regularization \cite{Carballo1,Carballo2}. Importantly, in 
their analysis they assume global hyperbolicity and, therefore, topology changes between spacelike slices are forbidden.  

In this work we study regular black holes from both a global, analytical and topological perspective with emphasis in the role of Borde's theorem and some its extensions. Here we will concern with singularity theorems, model geometries for regular black holes and topology change and causality violation within these systems. The work is organized as follows: section \ref{2} introduces preliminary definitions and some result from global techniques in order to make the work self-contained. ``Reversed" singularity theorems 
(here referred as {\it propositions}) {\it à la Borde} are formulated in Section \ref{Borde} trying to evade some classical singularity theorems in order to identify general features of regular black holes. After establishing possible model geometries for the core of spherically symmetric and (locally) static regular black holes in Section \ref{4}, we employ a topological approach based on Seifert bundles in order to track the topological transition between spacelike slices of most common regular black holes in Section \ref{5}, including a discussion of some issues such as the absence of global hyperbolicity and problems with causality. We end this section with a discussion on the advantages of regular black holes with a Nariai center. Final conclusions are left for section \ref{6}.

\section{Preliminary definitions}
\label{2}
The reader is referred to Refs. \cite{Hawkingbook,Oneill,Wald1984GeneralRelativity,Beem,Witten} for a detailed account of most of the definitions and properties here employed. However, 
here we include a brief summary with (hopely) all the necessary ingredients.
\\
\\
\textit{A spacetime} is a pair $(\mathcal{M},g)$ where $\mathcal{M}$ is a connected four-dimensional Hausdorff $C^{\infty}$ manifold and $g$ is a Lorentz metric on $\mathcal{M}$ (for brevity we will refer to $\mathcal{M}$ as a spacetime without explicit reference to the Lorentz metric).
\\
\\
The {\it chronological future (past)} of $p\in \mathcal{M}$, $I^{+(-)}(p)$ is the set of all $q\in \mathcal{M}$ such that there is a smooth future (past)-directed nondegenerate timelike curve from $p$ to $q$.
\\
\\
In case the preceding curve is causal (allowing the possibility of being degenerate) we define the {\it causal future (past)} of $p$, $J^{+(-)}(p)$.
\\
\\
A subset $S\subset\mathcal{M}$ of an arbitrary spacetime $\mathcal{M}$ is said to be \textit{achronal} if there does not exist a pair $p,q\in S$ such that it can be connected by causal curves. 
\\
\\
Let $\mathcal{S}$ be a spacelike three-manifold.
If every inextendible non-spacelike curve in $\mathcal{M}$ intersects $\mathcal{S}$, then $\mathcal{S}$ is said to be a {\it Cauchy surface}. A 
{\it partial Cauchy surface} is a closed achronal set S without edge (thus, a spacelike hypersurface).
\\
\\
$\mathcal{M}$ is said to be {\it globally hyperbolic} if it admits a global Cauchy surface. In this case, from Geroch's splitting theorem \cite{Geroch1970}, $\mathcal{M}$ is homeomorphic to $\mathbb{R}\times \mathcal{S}$. Even more, the extensions to the diffeomorphic have also been developed \cite{Bernal2003}.
\\
\\
A spacetime $\mathcal{M}$ is said \textit{future causally simple} if $E^+(X)=\dot I^+(X)$, where $\dot I^+(X)$ is the boundary of $I^+(X)$ and
$X$ is some compact achronal subset of $\mathcal{M}$. $E^+(X)$ is the future horismo of $X$, which is defined by $E^+(X)=J^+(X)-I^+(X)$.
\\
\\
\textit{A trapped surface} is a two-surface in which both outgoing and ingoing null geodesics perpendicular to this surface are convergent, {\it i. e.}, these null geodesics have negative divergence on this surface. 
\\
\\
For an \textit{eventually future-trapped surface} only is required that the divergences are negative somewhere in the future of the surface along each geodesic \cite{Borde1997RegularChange}.
\\
\\
A \textit{slice} $\Gamma$ is an edgeless, achronal hypersurface; {\it i. e.}, for every point $p\in\Gamma$ there is no timelike curve that can reach points $u\in I^-(p)$ and $v\in I^+(p)$. Even more, $\Gamma$ is a closed topological hypersurface \cite{Oneill}. In addition, if $\mathcal{M}$ is simply connected, then every closed spacelike hypersurface in $\mathcal{M}$ is achronal \cite{Oneill}.
\\
\\
The Ricci tensor obeys the {\it null curvature condition} (NCC) if $R_{\mu\nu}n^{\mu}n^{\nu}\geq0$ for all null vectors $n^{\mu}$ (we refer to {\it curvature} conditions when no specific dynamics, including Einstein's gravity, is assumed. Otherwise we will refer to {\it energy} conditions).
\\
\\
The {\it generic curvature condition} (GCC) says that every causal geodesic contains some point for which $k_{[\alpha} R _{\beta] \gamma \delta [\epsilon} k_{\phi]} k^{\gamma} k^{\delta}\ne 0 $, where $k_{\alpha}$ is tangent to the causal geodesic.
\\
\\
The {\it weak energy condtion} (WEC) is satisfied when $T_{\alpha\beta} t^{\alpha} t^{\beta} \ge 0$ for any timelike vector $t^{\alpha}$.
\\
\\
The {\it strong energy (curvature) condition} (SEC, SCC) is satisfied when $\left(T_{\alpha\beta}-\frac{1}{2}T g_{\alpha\beta} \right) t^{\alpha} t^{\beta} \ge 0$ 
($R_{\alpha \beta}t^{\alpha} t^{\beta} \ge 0$) for any timelike vector $t^{\alpha}$.
\\
\\
Finally, we say that a spacetime $\mathcal{M}$ is non-spacelike geodesically incomplete if $\mathcal{M}$ has a timelike or null geodesic which can not be defined for all values of an affine parameter. These spacetimes are said to be {\it singular}.
\\
\\
\indent
With these tools at hand, now we are ready to discuss some relevant results.

\section{Reversed singularity theorems}
\label{Borde}
{\it Theorem (Borde)}\, \cite{Borde1997RegularChange}. Suppose that there is a spacetime, $\mathcal{M}$, such that

\begin{enumerate}
    \item $\mathcal{M}$ contains an eventually future-trapped surface $\mathcal{T}$.
    \item The NCC is satisfied.
    \item $\mathcal{M}$ is null-geodesically complete to the future.
    \item $\mathcal{M}$ is future causally simple, {\it i. e.}, $E^+(X)=\dot I^+(X)$, where $X$ is any achronal compact subset of $\mathcal{M}$,
\end{enumerate}

then there is a compact slice to the causal future of $\mathcal{T}$.
\\
\\
\indent
{\it Sketch of the proof}. The key idea is to start from a ``reversed" Penrose's theorem \cite{Penrose1965} by assuming $\mathcal{M}$ to be geodesically complete to the future, together with a slightly different version of all the hypothesis involved in it with the exception on the existence of a non compact Cauchy surface. Then, Borde's theorem follows directly from this ``reversed" version (see Ref. \cite{Borde1997RegularChange} for details).
\\
\\
\indent
As commented in the Introduction, the seminal ideas of Sakharov \cite{Sakharov1966TheMatterb} and Gliner \cite{Gliner1966AlgebraicMatterb} have been widely used to substitute singularities by an inflationary equation of state. Usually, most of regular black hole models rely on spherical symmetry and isotropy for the core which, as exemplified by a de Sitter one, is tacitly assumed. Interestingly, Borde's theorem, which is usually taken as {\it the way} to evade Penrose's theorem (at least in the regular black holes literature), has nothing to say concerning the (an)isotropy of the core, which is usually identified with the compact slice referred to in the theorem. Therefore, we think it is of interest to look for regular black hole cores other that de Sitter but compatible with Borde's theorem without assuming isotropy. This point will be fully treated in Sect. \ref{4}. 
\\
\\
\indent
Even more, not only Borde's theorem can be used to avoid the formation of singulatities. One can enunciate several Borde's-like theorems employing other singularity theorems. For example, ``reversing" the famous Hawking and Penrose's singularity theorem \cite{HP1970} one obtains the following
\\
\\
\indent
{\it Proposition}. Suppose that there is a spacetime, $\mathcal{M}$, such that

\begin{enumerate}
    \item $\mathcal{M}$ contains a trapped surface or a compact spacelike surface or a point with a re-converging light cone.
    \item The GCC is satisfied.
    \item $\mathcal{M}$ is causally geodesically complete.
   \end{enumerate}

Then either $\mathcal{M}$ contains closed timelike curves or the SCC does not hold at some point (or both of them).
\\
\\
\indent 
{\it Comments on the proof(s)}. 
The proof of this and other propositions presented in the present work are a direct consequence of well-known singularity theorems. For example, if one of these theorems is generically expressed as $A_{1} \wedge A_{2}\rightarrow B$, where $A_{1,2}$ and $B$ are the assumptions and consequences of the theorem, respectively, and $\wedge$ stands for the logical ``and" symbol, the ``reversed" proposition we consider would be of the form $\lnot B \wedge A_{2} \rightarrow \lnot A_{1}$ ($\lnot$ stands for logical negation) and, therefore, their proof will follow immediately from that of the theorem previously stated. Thus, in what follows, although no specific proofs will be explicitly presented for the rest of the propositions, their validity is logically guaranteed.  Although Borde's theorem is, perhaps, the best-known example of the previous strategy, here we will systematically employ it in order to prove some interesting results concerning the interplay between regularity, topology and causality.
\\
\\
The physical consequences of the previous proposition can be undestood, for instance, by looking at the spherically symmetric and static case. In this case, Mars, Martín-Prats and Senovilla proved that if these spacetimes are regular at $r=0$ and satisfy $\rho + p_{r}+2 p_{t}\ge 0$, which is a consequence of the SEC, then they cannot contain any black hole region in General Relativity \cite{Mars1996}. Therefore, in this precise sense, we can assure that 
by reversing Hawking and Penrose's theorem one can conclude that 
``regular black holes violate the strong energy condition". Here we note that, although there are particular models where
explicit calculations have shown that the SEC is violated in regular black holes \cite{Elizalde2002}, an explicit reference to the Hawking and Penrose's theorem has been only found in Ref. \cite{Zasla2010}. Interestingly, the violation of the SEC inside the event horizon gives place to a negative Tolman mass \cite{Zasla2010}, which changes its sign inside the Cauchy horizon depending on the singular or regular character of the black hole. Even more, it has been recently conjectured that these features could be related to topology changes in regular black holes models \cite{Melgarejo2020}, which we will discuss in Sect. \ref{5}.
\\
 \\
\indent
With respect to the GCC, some comments are in order: (i) it represents almost no restriction on generic spacetimes \cite{Hawkingbook,Beem,HP1970}; (ii) it can be violated in some spacetimes specialized from the geometrical point of view (e. g., Reissner-Nordstr\"om, as pointed out in \cite{Borde1994}); (iii) the strict SCC implies the GCC (see propositions 2.5 and 2.6 of \cite{Senovilla1997}). Interestingly, this last point leaves room for the following 
result, which is also a direct consequence of the Hawking-Penrose theorem \cite{HP1970}:
\\
\\
\indent
{\it Proposition}. Suppose that there is a spacetime, $\mathcal{M}$, such that

\begin{enumerate}
    \item $\mathcal{M}$ contains a trapped surface or a compact spacelike surface or a point with a re-converging light cone.
    \item The SCC is satisfied.
    \item $\mathcal{M}$ is causally geodesically complete.
    \item  $\mathcal{M}$ does not contain closed timelike curves.
   \end{enumerate}

Then the GCC does not hold at some point of some causal geodesic of $\mathcal{M}$.
\\
\\
\indent
From this result we can extract an important conclusion: in general (for example, for non-spherically symmetric spacetimes), the violation of the SCC (of SEC) is not mandatory for regular black holes (wrong assertions regarding this point are frequent in the literature, for example in \cite{Eliasplb} and \cite{Rodrigues2016}). Of course, this is at the price of having $R_{\alpha \beta} t^{\alpha} t^{\beta} = 0$ and 
$R_{\alpha \beta}n^{\alpha} n^{\beta} = 0$ in the timelike and null cases, respectively \cite{Beem}, which is a consequence of the violation of the GCC at some causal geodesic. In fact, a violation of at least some of the {\it curvature} (not {\it energy)} conditions (including the generic one) must occur if a regular solution must be present. This is a subtle point which we have not been able to find in the literature. Even more, in the spherically symmetric case, the violation of the GCC on some radial null geodesic implies the Schwarzschild ansatz, $g^{tt}g_{rr}=-1$ \cite{Jacobson2007}, which is used, up to our knowledge, in all spherically symmetric regular black hole solutions. It is important to note that the opposite is not true; {\it i. e.}, the Schwarzschild ansatz does not imply that the GCC hods. This can be see, for example, in the Reissner-Nordstr\"om solution, as previously commented.
\\
\\
\indent
From a physical point of view, it would be desirable to have black holes with nice properties such as, for example, properties 1-4 of the previous proposition. This motivates the following
\\
\\
\indent {\it Definition.} A black hole is {\it regular and well-behaved} if  (1)-(4) of the previous proposition are satisfied.
\\
\\
\indent
Even more, based on the previous proposition, an interesting conclusion can be reached in the spherically symmetric case considered in Ref. \cite{Mars1996} in the following sense:
\\
\\
\indent
{\it Proposition}. Let $\mathcal{M}$ be a spacetime containing a regular and well-behaved black hole. Then, the corresponding theory is not General Relativity and either (i) the SEC is saturated at some point along some timelike geodesic of $\mathcal{M}$ or (ii) the NEC is saturated at some point along some null geodesic of $\mathcal{M}$.
\\
\\
\indent
As a simple application of this result we note that, for a perfect fluid, the SEC can not be saturated at any point (for example, in the isotropic case, $\rho + p=0$ and $\rho+3 p=0$ can not hold simultaneously for a matter content other than vacuum). Therefore, the only possibility is that the NEC is saturated at some point with implies that the geometry under consideration is de-Sitter like at this particular point. Of course this does not imply that the core of these objects has to be describe by a de-Sitter geometry but it excludes the possibility of having regular and well behaved black holes beyond General relativity if the geometry is de Sitter nowhere.
\\
\\
\indent
Although the previous results show some extra properties that regular and well-behaved black holes have to fulfill, including 
specific ways of evading Hawking and Penrose's theorem and their consequences for the SEC and for the GCC in regular black holes, in the next section we will focus on some consequences of Borde's theorem, which, although it implies that the SEC is violated somewhere, it is usually referred to in regular black holes literature.

\section{Model geometries for spacelike slices}
\label{4}
In general, four dimensional non rotating electrovacuum black holes have a topology given by $\mathbb{R}^2\times \Sigma$, where $\Sigma$ is any closed surface of constant curvature and arbitrary genus, $g$ \cite{Vanzo1997}. For simplicity, only the $g=0$ case will be considered here. The case of a non-vanishing cosmological constant will be treated separately.

\subsection{Analytical approach}

\subsubsection{(Anti-)de Sitter cores}

Let us introduce a coordinate system in any spacelike slice of these spacetimes such that the line element can be written as
\begin{equation}
	\label{3metric}
	ds^2= \frac{r^2}{\lambda(r)}dr^2+ r^2\left(d\theta^2 + f(\theta) d\phi^2 \right).
\end{equation}
\\
\indent
On one hand, an straightforward computation reveals that the scalar curvature is given by
\begin{equation}
	R(r,\theta)= \frac{2 \lambda}{r^4}+\frac{\dot f}{2 f^2 r^2}-\frac{2 \lambda'}{r^3}-\frac{\ddot f}{r^2 f},
\end{equation}
\\
where $\lambda'\equiv \frac{d \lambda}{dr}$  $\dot f \equiv \frac{d f}{d \theta}$. The formal solution is given by
\begin{eqnarray}
\label{lambda}
	\lambda(r)&=& r A_{1} + r \int_{1}^{r}\frac{\left(\dot f\right)^2-2 f\left(f y ^2 R(y,\theta)+\ddot f\right)}{4 f^2}dy,
\end{eqnarray}
\\
where $A_{1}$ is an arbitrary constant.
\\
\\
\indent
Interestingly, Eq. (\ref{lambda}) can be expressed for $r\rightarrow 0$ as

\begin{eqnarray}
    \lambda(r)&\simeq & r \bigg(A_{1}+\int_{1}^{0}\frac{\left(\dot f\right)^2-2 f\left(f y ^2 R(y,\theta)+\ddot f\right)}{4 f^2}dy \bigg) \nonumber \\
    &+& \frac{r^2}{4}\bigg[\left(\frac{\dot f}{f}\right)^2- \frac{2 \ddot f}{f} \bigg] -\frac{1}{6}R(0,\theta) r^4,
\end{eqnarray}
\\
\indent
which gives place to

\begin{equation}
    R(r,\theta) \simeq R(0,\theta),
\end{equation}
\\
for $r\rightarrow 0$.
\\
\\
\indent
Now let us focus our attention on the second term of the rhs of this series expansion. It is clear that it must be constant in order for $\lambda$ and $R$ to be functions only of the radial coordinate. In this case we have
\begin{equation}
\label{f}
    \left(\frac{\dot f}{f}\right)^2- \frac{2 \ddot f}{f} = k,
\end{equation}
\\
\indent
where we have chosen $k=0,\pm 1$ (this choice will be later understood). With an appropriate
choice for the integration constants, the solutions of
Eq. (\ref{f}) are:
\begin{eqnarray}
    k = &1& \, \, \,f(\theta) = \sin^2\theta \nonumber \\
    k =-&1& \, \, f(\theta) = \sinh^2\theta \nonumber \\
    k = &0& \, \, f(\theta) = 1 .
\end{eqnarray}
\\
\indent
On the other hand, the Kretschmann scalar, $\mathcal{K}=R^{\alpha\beta\gamma\delta}R_{\alpha\beta\gamma\delta}$, is given at $r\rightarrow 0$ by
\begin{equation}
    \mathcal{K}\simeq\frac{6}{r^6}\left(\int_1^0 \left(k-\frac{1}{2}y^2 R(y)\right) \, d y+  A\right)^2, 
\end{equation}
\\
where $A$ is an arbitrary constant.
\\
\\
\indent
Let us impose $\lim_{r\rightarrow 0}\mathcal{K}\rightarrow $ finite (the same reasoning applies for 
$R^{\alpha \beta}R_{\alpha \beta}$). Then, we have to choose 
$A=-\int_1^0 \left(k-\frac{1}{2} y^2 R(y)\right) dy$. In this case, we get

\begin{equation}
\label{eqlambda}
	\lambda(r)\simeq k\, r^2-\frac{1}{6}R(0)r^4.
\end{equation}
\\
\indent
Now let us introduce an angular coordinate, $\chi$, such that 
\begin{eqnarray}
	1-\frac{r^2 R(0)}{6}&\equiv&\cos^{2} \chi \, \, (k=1) \nonumber \\
	1+\frac{r^2 R(0)}{6}&\equiv&\cosh^{2} \chi \, \, (k=-1).
\end{eqnarray}
\\
\indent
Within these new coordinates, Eq. (\ref{3metric}) now reads
\begin{eqnarray}
	ds^2 &=& \frac{6}{R(0)}\left(d\chi^2+\sin^2\chi \left(d\theta^2 +\sin^2\theta\, d\phi^2  \right) \right) \nonumber \\
	ds^2&=&\frac{6}{R(0)}\left(d\chi^2+\sin^2\chi \left(d\theta^2+\sinh^2 \theta \,d\phi^2  \right) \right) 
\end{eqnarray}
\\
and, therefore, the topology of the spacelike slices are described by either 
$S^3$ (corresponding to a de Sitter core $R(0)>0$) or $H^3$ (corresponding to an anti-de Sitter core, $R(0)<0$). The $k=0$ case, although slightly more elaborated, gives place to a three-torus \cite{Vanzo1997} . This case will be considered elsewhere.
\\
\\
\indent
Interestingly, in the $\Sigma=H^2$ case, although the metric displays the properties of a black hole, it is not in fact, as it represents the portion of AdS which is causally accessible to a family of accelerated observers \cite{Vanzo1997}. In addition,
$H^2$ is non compact. However, we can make the quotient $H^2/G$ with an appropriate discrete group in order to make the
horizon a compact Riemann surface of genus $g>1$ \cite{Vanzo1997}.
\\
\\
\indent
Concerning the whole spacetime but not only a spatial slice, it has been shown (see, for example, Ref. \cite{Garcia2019} and references therein) that spherically symmetric and static regular black hole solutions must approach to a de-Sitter spacetime at the regular center. In addition, regularity conditions have been studied for rotating black holes\cite{Torres2017}. Finally, the case for planar and cylindrical regular spacetimes, although considered some years ago \cite{Lake1994}, has not received too much attention. 
\\
\\
\indent
We end this section by noting that a careful inspection of Eq. (\ref{eqlambda}) reveals that, in case $R(0)=0$, the so-called Simpson and Visser's {\it hollow} regular
black holes \cite{Visser2019} appear. Interestingly, although this family of regular black holes has not been explored in depth, it has some desirable
properties. For example, as the usual de Sitter core is substituted by a Minkowskian one, there are no topology changes between slices. Even more, by a direct application of Borde's theorem we infer that these black holes must violate the NEC (the rest of hypothesis of Borde's theorem are satisfied), in agreement with recent calculations \cite{Visser2019}.
\\
\\
\indent The same authors have recently introduced \cite{Visser2019bis} a parametric family of spherically symmetric geometries with interpolate between the Schwarzschild black hole and a traversable wormhole through a regular black hole. Specifically, the line element reads
\\
\begin{equation}
\label{metricvisser}
    ds^2=-\left(1-\frac{2m}{\sqrt{r^2+a^2}}\right)dt^2+\frac{dr^2}{1-\frac{2m}{\sqrt{r^2+a^2}}}+\left(r^2+a^2\right)d\Omega^2.
\end{equation}
\\
\indent
 It represents a regular black hole when $a\in (0,2m)$ and $m>0$. Interestingly, the interpretation of this geometry near the regular center can be easily read from the components of the corresponding energy-momentum tensor. Using Eq. (5.1) of \cite{Visser2019bis} we get that both $\rho+p_{\bot}\ne 0$ and $\rho+p_{\parallel}\ne 0$ near the regular center within the previous range of values for $a$. Therefore, we are again with a regular black hole solution without a de Sitter center. As in the case of {\it hollow} regular black hole, the NEC is again violated but, in this case, Borde's theorem does not apply. 
 \\
 \\
 \indent
Even more, near the regular center and for $a \in (0,2m)$, Eq. (\ref{metricvisser}) reads
\begin{equation}
\label{nariailike}
    ds^2 \approx \left(|\epsilon|-\frac{1+|\epsilon|}{2 a^2}r^2 \right)dt^2-\frac{dr^2}{|\epsilon|-\frac{1+|\epsilon|}{2 a^2}r^2 } + a^2 d\Omega^2,
\end{equation}
where $|\epsilon|=\frac{2m}{a}-1$. Interestingly, this geometry can be interpreted as AdS$_{2}\times S^2$ (note that this geometry is not of the form of Eq. (\ref{3metric})). In general, these kind of geometries such as Eq. (\ref{nariailike}) belong to the family of regular black holes with a Bertotti-Robinson core, as we will show here.

\subsubsection{Nariai and Bertotti-Robinson cores}

In this case we start from the spherically symmetric spacetime whose line element is given by
\begin{equation}
\label{metric}
    ds^2 = -\frac{\lambda(r)}{r^2}dt^2+\frac{r^2}{\lambda(r)}dr^2+l^2 d\Omega^2,
    \end{equation}
\\
\\
where $l$ is a parameter with dimensions of length which can be related either to the cosmological constant or to other relevant length scale present in the problem under consideration.
\\
\\
\indent An straightforward calculation reveals that
\begin{widetext}
\begin{equation}
    \lambda(r)=r^2\Bigg[C_{1}+ C_{2} r+ \int_{1}^{r}y(r)\left(R(y)-\frac{2}{l^2}\right)dy +r \int_{1}^{r}\left(R(y)-\frac{2}{l^2}\right) dy\Bigg],
\end{equation}
    \end{widetext}
where $C_{1,2}$ are arbitrary constants.
\\
\\
\indent
Therefore, provided $R(r)$ is regular everywhere, we have that
\begin{eqnarray}
\mathrm{Ric}^2 &=& \frac{4}{l^4}+\frac{R}{2}\left(R-\frac{4}{l^2} \right) \nonumber \\
\mathcal{K}&=& \frac{8}{l^4}+R\left(R-\frac{4}{l^2}\right)\\
\end{eqnarray}
\\
are regular too.
\\
\\
\indent Even more, if the integration constants are chosen such that
\begin{eqnarray}
C_{1}&=&1-\int_{1}^{0}y(r)\left(R(y)-\frac{2}{l^2}\right)dy \nonumber \\
C_{2}&=&-\int_{1}^{0}\left(R(y)-\frac{2}{l^2}\right)dy
\end{eqnarray}
\\
\indent then we have that, near the $r=0$ regular center,
\begin{equation}
    \lambda(r)\simeq r ^2+\left(\frac{1}{l^2}-\frac{R(0)}{2}\right)r^4
\end{equation}
\\
and, therefore, Eq. (\ref{metric}) reads, near the core,
\begin{equation}
\label{core}
    ds^2 \simeq -\left[1+\left(\frac{1}{l^2}-\frac{R(0)}{2}\right)r^2\right] dt^2+\frac{dr^2}{1+\left(\frac{1}{l^2}-\frac{R(0)}{2}\right)r^2} +l^2 d\Omega^2.
    \end{equation}
%
\\
\\
\indent The appearance of these geometries makes the NEC not to be saturated near the core, as it can be seen using Einstein's equations. Defining the effective component of the energy-momentum tensor as usual, we get ($8 \pi G =1$)
\begin{eqnarray}
\rho(r) &=&\frac{1}{l^2} \nonumber \\
p_{\parallel}(r) &= &-\frac{1}{l^2} \nonumber \\
p_{\bot}(r) &= &\frac{6 \lambda - 4 r \lambda'+ r^2 \lambda ''}{2 r^4}
\end{eqnarray}
\\
and, therefore, $\rho+p_{\parallel}=0$ but $\rho+p_{\bot}\ne0$.
\\
\\\indent Clearly, Eq. (\ref{core}) is $\mathrm{dS}_2\times S^2$ when $\frac{1}{l^2}-\frac{R(0)}{2}<0$ and $\mathrm{AdS}_2\times S^2$ when $\frac{1}{l^2}-\frac{R(0)}{2}>0$. Let us take a closer look at these geometries.
\\
\\
\indent
The neutral Nariai solution, first introduced by Nariai \cite{Nariai1}, solves Einstein's
field equations with a positive cosmological constant and is given by
\begin{equation}
\label{Nariai}
    ds^2=\Lambda^{-1}\Bigg(-\sin^2\chi\, dt^2+d\chi^2+d\theta^2+\sin^2 \theta \, d\phi^2 \Bigg),
\end{equation}
\\
where $\chi$ and $\theta$ both run from 0 to $\pi$, and $\phi$ has period $2\pi$. Importantly, the spacelike slices are $S^1\times S^2$ in this case (the same occurs for the charged Nariai solution, found by Bertotti and Robinson \cite{BR1}). Essentially, it suffices to make the coordinate change
$1-|\frac{1}{l^2}-\frac{R(0)}{2}|r^2\equiv \cos^2\chi$ to bring Eq. (\ref{core}) to the form of Eq. (\ref{Nariai}).

\indent The simplest Bertotti-Robinson solution \cite{BR1} solves the $\Lambda=0$ Einstein-Maxwell system. The line element is given by
\begin{equation}
\label{BR2}
    ds^2=q^2\Bigg(-\sinh^2\chi\, dt^2+d\chi^2+d\theta^2+\sin^2 \theta \, d\phi^2 \Bigg),
\end{equation}
\\
where $q$ is the charge of the solution, $\chi$ is unbounded, $\theta$ runs from 0 to $\pi$, and $\phi$ has period $2\pi$. In this case, the spacelike slices are $\mathbb{R}\times S^2$. A redefinition of the form $\sinh^2 \chi \equiv \frac{R^2}{q^2}-1$ and $t=\frac{T}{q}$ brings Eq. (\ref{BR2}) to 
\begin{equation}
\label{BR}
    ds^2=-\Bigg(\frac{R^2}{q^2}-1\Bigg) dT^2+\frac{dR^2}{\frac{R^2}{q^2}-1}+
    q^2\Bigg(d\theta^2+\sin^2\theta d\phi^2\Bigg),
\end{equation}
\\
which is $\mathrm{AdS}_{2}\times S^2$ written in Poincaré-like coordinates \cite{Natsuume2015}.
\\
\\
\indent
Summarizing, we have found that the geometry of spacelike slices of spherically symmetric and (locally) static regular solutions is:
\begin{itemize}
    \item $S^3$ for a dS core
    \item $H^3$ for an AdS core
    \item $S^1\times S^2$ for a Nariai core
    \item $\mathbb{R}\times S^2$ for a Bertotti-Robinson core
\end{itemize}

\indent
Although both the AdS and the Bertotti-Robinson cores are not compact and, therefore, Borde's theorem implies that the NCC is not satisfied (assuming causal simplicity), these geometries turn to be fundamental in order to avoid some causality issues (see section V B).

\subsection{Topological approach}

Let us require that the spacelike slices, $\Gamma$, are: (i) smooth and (ii) simply connected. As previously stated, $\Gamma$ is achronal and edgeless and, therefore, it acquires the structure of a topological manifold. The extension to $C^\infty$ class is taken as a physical requirement. With these extra ingredients we are ready to formulate the following corollary to Borde's theorem. 
\\
\\
{\it Corollary}. Assume $\Gamma$ smooth and simply connected. Then, $\Gamma \simeq S^3$.
\\
\\
{\it Proof}. Direct from the Poincar\'e conjecture \cite{Ricciflow}.
\\
\\
\indent 
Therefore, if $\Gamma$ is smooth but not $S^3$, then $\Gamma$ is not simply connected. Thus, slices with, for example, $S^1\times S^2$ and $S^1\tilde \times S^2$ (the non-orientable circle bundle over $S^2$) topologies are compatible with Borde's theorem. This can be seen for the specific case of regular black holes with their Carter-Penrose diagram {\it à la} Reissner-Nordstr\"om by carefully performing null identifications \cite{Melgarejo2020} and for the Nariai family, previously presented.

\section{Topology change in regular black holes}
\label{5}
We have seen that the class of regular black holes satisfying Borde's theorem  have the topology of $S^3$ at their cores if the compact slice is assumed to be smooth and simply connected. As commented in the introduction, all regular black hole solutions studied (with few exceptions) belong to this class. Therefore, as the ``slice at infinity" is usually considered to be $\Gamma_1 =\mathbb{R}\times\Sigma$ (with $\Sigma=S^2$ is most cases), there is a transition between
$\Gamma_1$ and $\Gamma_2=S^3$ slices. It is our purpose to describe and quantify this transition with the help of Seifert bundles. We refer the interested reader to \cite{Orlik1972} for details on Seifert manifolds.

\subsection{Seifert bundles}
The clearest definition of a {\it Seifert fiber space} is a three-manifold which can be foliated by circles. However, it is usually more useful to think of it as a kind of bundle over a two-dimensional manifold (or over a two-dimensional orbifold, in general) with fibre the circle.

If the manifold is compact, foliation by circles is usually more obvious that for the non-compact case. For example, in the $\Gamma=S^3$ case, which corresponds to the de Sitter core, the foliation corresponds to the well-known Hopf fibration. In the non-compact case we can mention, following Ref. \cite{Scott1983}, that any 3-manifold with a geometric structure modelled on $\mathbb{R}\times S^2$ is foliated by lines or circles. Therefore, the slices we are referring to in this work are Seifert fiber spaces. Using some tools from these spaces, it is our purpose to show here how to distinguish between the core and the rest of the slices. 

In the two dimensional case, the geometry of a given closed surface $\Sigma$ is given by the Euler number, $\chi (\Sigma)$, and whether $\chi$ is positive, zero or negative. When dealing with Seifert bundles, the appropriate geometry can be determined from two invariants: the Euler number of the base manifold (or, in general, of the base orbifold), $\chi$, and the Euler number of the Seifert bundle, $e$. For the
case we are interested in we have \cite{Scott1983}:
\begin{eqnarray}
    \Gamma_{1}=\mathbb{R}\times S^2: &&\, \, \, \chi =2  \, \, \, \mathrm{and}\, \, \, e = 0 \nonumber \\
    \Gamma_{2}=S^3: &&\, \, \, \chi =2  \, \, \,\mathrm{and}\, \, \,e =1 \nonumber 
\end{eqnarray}
\\
\indent
Note that, although the base manifolds are $S^2$ and, therefore, $\chi=2$, differences between $\Gamma_1$ and $\Gamma_2$ are due to the triviality of the bundle.
\\
\\
\indent
First of all, note that there is a direct way of computing the Euler number of the Seifert bundle which is associated to the corresponding Seifert invariant:
\begin{equation}
\label{edef}
    e= -\sum_{j=1}^{n}\frac{\beta_{j}}{\alpha_{j}}, 
\end{equation}
\\
where the relative prime pairs, $(\alpha_{j},\beta_{j})$, determines the orbit type of an orbit with isotropy group $\mathbb{Z}_{\alpha_{j}}$ \cite{Orlik1972}. From the previous definition it can be
seen that $e\left(\Gamma_{1}\right)=0$ but $e\left(\Gamma_{2}\right)=1$. 
\\
\\
\indent
In addition to this way of calculating $e$, there is a nice result which could be useful in the physics literature. Here we will briefly summarize the construction developed in Ref. 
\cite{Ouyang1994} without giving any proofs, which can be found in the aforementioned work. From this point we will refer only to manifolds and not to orbifolds, although appropriate generalizations can be usually performed.
\\
\\
\indent
Let $\Sigma$ be a closed two-dimensional Riemannian manifold. For the special case of constant Gauss curvature of $\Sigma$, $K$, the Gauss-Bonnet theorem states that
\begin{equation}
\label{gauss}
    K \,\mathrm{Vol(\Sigma)= 2\pi \chi (\Sigma)}.
\end{equation}
\\
\indent
Interestingly, there is an expression similar to Eq. (\ref{gauss}) but for the case of certain circle bundles over Riemannian manifold, which reads 
\begin{equation}
\label{gaussbis}
\tilde R \, \mathrm{Vol(B)}= 2\pi e (\Gamma).     
\end{equation}
\\
\indent
In Eq. (\ref{gaussbis}), $\tilde R$ is the (not necessarily constant) Riemann curvature of the fiber metric $\tilde g$ in the total space, $\Gamma$, and $\mathrm{Vol(B)}$ is the volume of the base manifold.
\\
\\
\indent
On one hand, from Eq. (\ref{edef}) we get that $\tilde R$ is constant on $\Gamma_1$, which can be shown by direct calculation. In fact, $\tilde R = 0$ for $\Gamma_{1}$ and $\tilde R = 2 K$ for $\Gamma_{2}$, where $K$ is the constant Gaussian curvature of $\mathrm{B}$, which is a two-sphere. Note that the factor of $2$ is irrelevant since it can be absorbed in the normalization of $e$. Then, Eqs. (\ref{gauss}) and (\ref{gaussbis}) are completely equivalent with the exception that $\tilde R$ contains information of both the fibre and the base. 
\\
\\
\indent
Therefore, this curvature, $\tilde R$, can be used to distinguish between slices of singular and regular (in the sense of Borde's theorem) black holes, as summarized in the following table
\\
\begin{table}[htb]
\centering
\begin{tabular}{ |c|c|c|c| } 
 \hline
 Slice & $\chi(B)$ &$e(\Gamma)$ &  $\tilde R$ \\ 
 \hline
 $\mathbb{R}\times S^2$& 2 & 0&0\\ 
 $S^3$& 2 &  1&$2 K(\mathrm{B})$ \\
 \hline
 \end{tabular}
 \caption{Differences between achronal slices of spherically symmetric regular black hole spacetimes.}
 \label{table}
\end{table}

\subsection{Some issues: predictability and causality violation}

Interestingly, by Geroch's splitting theorem \cite{Geroch1970}, these regular black hole spacetimes are not globally hyperbolic. Although this might be though in principle as a major drawback associated with lack of predictability, Wald showed \cite{W1} that, for the case of a Klein-Gordon
scalar field propagating in an arbitrary static space-time, a physically well-posed and fully deterministic dynamical evolution prescription can be given. Even more, this prescription
is the only possible way of defining the dynamics of a scalar field in
a static, non-globally-hyperbolic, spacetime, as shown by Wald and Ishibashi \cite{W2}, who applied their techniques to the dynamics of electromagnetic and gravitational perturbations in an AdS spacetime, which is a widely used example of the lack of global hyperbolicity \cite{W3}.  Therefore, as the kind of regular black holes we are referring to in this work are static, Wald and Ishibashi's prescription can be safely employed to restore determinism.
\\
\\
\indent
Determinism is not the only problem faced by regular black holes. Topology change, which seems to be unavoidable within these systems, is usually believed to occur at a high price of causality violations due to results by Geroch \cite{Geroch1,Geroch2}. Even more, Tipler showed that Einstein’s equations cannot hold (with a source with non-negative energy density) if the spatial topology changes \cite{Tipler1,Tipler2}. One of the main assumptions of these and other theorems are the compactness of the spatial slices. Therefore, as one situation of interest is the asymptotically flat spatial geometry of
an isolated system (including regular black holes), these results have to be extended. As pointed out by Borde \cite{Bordetopologychange} ``We expect to be able to compactify this situation by adding a point at infinity, or by imposing periodic boundary conditions (“putting it in
a box”), and thus we might expect results similar to the compact case." In this line of thought, Geroch introduced the concept of externally Euclidean three-manifold \cite{Geroch2}, which can represent an isolated system, to show that the topology change occurs within certain compact set. 
\\
\\
\indent
Concerning cobordism and Morse theory, most of the techniques refer to compact slices \cite{Antonelli1979,Sorkin1986, Horowitz1991,Gibbons1992,Low1992}. In fact, up to my knowledge, only few works mention how to implement cobordism theory for non-compact slices. Specifically, Yodzis comments \cite{Yodzis1972,Yodzis1973} that the standard results could be used when two slices are related by a finite number of surgeries (which is always possible if both are compact), although no specific examples are given. Dowker and García have developed \cite{Fowker1998} a Morse and handlebody technology that can be adapted to non-compact manifolds. As the authors said: ``This will not be difficult because, with the assumption that the
topology change is localized in space, we can reduce the questions to the closed case by, roughly speaking, closing off space [...] (and then) we open back to the physical manifolds." 
\\
\\
\indent
Therefore, although most of these techniques are, in principle, applicable to the case of regular black holes here considered and they could serve to shed more light on the relation between regularity and topology change on general grounds, here we will briefly focus on Geroch's extensions of his classical theorems \cite{Geroch1,Geroch2} to the case of open slices (see also \cite{Bordetopologychange} and \cite{Tipler2} for a brief account of these techniques).

\subsection{Geroch and Tipler's theorems and regular black holes}

As previously mentioned, regular (and most of singular) black holes usually have open slices with $\mathbb{R}\times S^2$ topology. Therefore, some ``compactness requirement" has to be introduced in order to apply the usual techniques. This is captured by the following
\\
\\
\indent
{\it Definition \cite{Geroch2}}. A three-manifold $\Gamma$ is said to be {\it externally Euclidean} if there exists a connected compact set $C$ of $\Gamma$ such that $\Gamma-C$ is diffeomorphic to $\mathbb{R}\times S^2$ (this means that $\Gamma-C$ is diffeomorphic to the Euclidean space minus a three-ball).
\\
\\
\indent
{\it Definition \cite{Geroch2}}. Let $\mathbb{M}$ be a four-dimensional subset of a spacetime, $\mathcal{M}$, whose boundary is the disjoint union of two three-manifolds $\Gamma$ and $\Gamma'$ which are externally Euclidean and spacelike. Suppose that there exists a connected compact set $K$ of $\mathbb{M}$ such that $\mathbb{M} - K$ is diffeomorphic to $S^2\times \mathbb{R}\times [0,1]$, where for each fixed number $\alpha \in [0,1]$ the submanifold $S^2\times \mathbb{R}$ of $\mathbb{M}$ is spacelike and for each fixed point $p$ of $S^2\times \mathbb{R}$ the line $[0,1]$ of $\mathbb{M}$ is timelike. Then 
$\mathbb{M}$ will be called {\it externally Lorentzian}. 
\\
\\
\indent
Then, topology change (if any) must take place in the timelike world-tube, $K$, between $\Gamma$ and $\Gamma'$. 
\\
\\
\indent
With these definitions we are ready to state the fundamental result by Geroch. 
\\
\\
\indent{\it Theorem (Geroch) \cite{Geroch2}}. Let $\mathbb{M}$ be an externally Lorentzian portion of the spacetime, $\mathcal{M}$, the boundary of $\mathbb{M}$ being the disjoint union of two spacelike externally
Euclidean 3-manifolds, $\Gamma$ and $\Gamma'$. Suppose $\mathbb{M}$ has no closed timelike curves. Then $\Gamma\simeq \Gamma'$ and $\Gamma'$ and $\mathbb{M}\simeq \Gamma \times [0,1]$.

In case of the kind of regular black holes here considered, $\Gamma = \mathbb{R}\times S^2 \not \simeq \Gamma'= S^3$, with $\Gamma$ and $\Gamma'$ being externally Euclidean and compact, respectively. Therefore, Geroch's theorem applies and, therefore, there exist closed timelike curves within the timelike tube, $K$. Therefore, 
by ``reversing" Geroch's theorem we can state the following
\\
\\
\indent {\it Proposition}. There is no spacetime $\mathcal{M}$ containing regular and well behaved black holes. The existence of a de Sitter core and an asymptotically flat region implies the existence of closed timelike curves.
\\
\\
\indent Given this impossibility, it would be interesting to find regular black hole solutions which minimize the causality issue. As the problematic region is the timelike tube, $K$, one could ask for the smallest volume contained within $K$, which implies minimizing the surface area $\mathcal{A} \sim R(0)^{-3/2} $ of $\Gamma=S^3$. If we assume $R(0)\sim l_{p}^{-2}$ ($l_{p}$ stands for the Planck length) in the region near the core, where quantum gravitational effects are expected to occur, we get $\mathcal{A}\sim l_{p}^3$ near the core. Then, if a sufficiently small $C$ is assumed for the second slice $\Gamma'$ in a neighbourhood of the core, causality violation can be constrained to the Planck scale.
\\
\\
\indent Interestingly, Geroch's previous result was strengthened by Tipler, who proved the following
\\
\\
\indent 
{\it Theorem (Tipler) \cite{Tipler2}}.  Let $\mathbb{M}$ be an externally Lorentzian portion of an spacetime, $\mathcal{M}$, the boundary of $\mathbb{M}$ being the disjoint union of two spacelike externally
Euclidean 3-manifolds, $\Gamma$ and $\Gamma'$. If $\Gamma$ is a partial Cauchy surface for the entire spacetime, and a Cauchy surface
for $\mathbb{M}- K$, and in addition we assume (i) the WEC and the Einstein equations hold on $\mathbb{M}$; (ii) the GCC holds on $\mathcal{M}$.
Then $\Gamma\simeq \Gamma'$ and $\Gamma$ is a Cauchy surface for $\mathbb{M}$.
\\
\\
\indent In this case, our application to regular black holes can be stated as the following proposition, which is a direct consequence of ``reversing" Tipler's theorem.
\\
\\
\indent{\it Proposition}. Changes in the topology of spacelike slices of a regular black hole spacetime are not compatible with both the WEC and Einstein's equations and the GCC simultaneously.
\\
\\
\indent Therefore, assuming the GCC, going beyond General Relativity is revealed as a possibility of constructing regular black holes with topology change. In this case, the GCC is fulfilled but, as we previously said, the existence of closed timelike curves is unavoidable. 

\subsection{Regular black holes without topology change?}

At this point it should be clear that the compactness hypothesis for the Cauchy surface of Penrose's theorem \cite{Penrose1965} is the reason behind topology changes in regular black holes. Therefore, in order to avoid causality problems, regular black holes without topology changes have to be considered. 
\\
\\
\indent
Fortunately, there are singularity theorems for ``open" universes so that we can ``reverse" them to obtain interesting results from regular black holes. Incidentally (or not), one of these theorems was formulated by Borde and Vilenkin as follows: 
\\
\\
\indent {\it Theorem (Borde and Vilenkin) \cite{BordeVilenkin}}. Let $\mathcal{M}$ be a spacetime such that
\begin{enumerate}
    \item The NCC holds.
    \item $\mathcal{M}$ is future causally simple.
    \item $\mathcal{M}$ contains no compact slices.
    \item $\mathcal{M}$ contains a point $p$ whose future light cone reconverges.
\end{enumerate}

Then $\mathcal{M}$ is null geodesically incomplete to the future.
\\
\\
\indent Roughly speaking, the causally simple assumption is desirable. Although this condition is weaker than global hyperbolicity (see \cite{Minguzzi2008} for an up to date complete causal hierarchy), it suffices to assure, for example, that no closed causal curves are present.
\\
\\
\indent
Interestingly, the assumptions (1)-(4) of the previous theorem can be used to improve the previous definition of a regular and well-behaved black hole. Here we introduce the following
\\
\\
{\it Definition.} A spacetime $\mathcal{M}$ is said to contain an {\it ideal black hole} if assumptions (1)-(4) of the previous theorem are fullfilled.
\\
\\
\indent
Therefore, from the previous theorem we can state the following
\\
\\
\indent{\it Proposition}. No spacetime $\mathcal{M}$ can contain an ideal black hole.
\\
\\
\indent Thus, regular black holes without either topology changes or closed timelike curves and satisfying the NCC are not supported.
\\
\\
\indent At this point, some comments are in order: (i) Assumptions (2) and (3) can not be relaxed if we want that $\mathcal{M}$ does not contain closed timelike curves; (ii) assumption (3) introduces one of the main features of black holes (equivalent to the existence of trapped surfaces) and, therefore, it can not be deleted. Then, we arrive to the conclusion that only the NCC can be relaxed in order to obtain a closer version of an {\it ideal black hole}. Perhaps instead of introducing a de Sitter-like core (say $\rho + p=0$ to simplify) to regularize the $r=0$ region of singular black holes within spherical symmetry, one could replace it by an effective fluid minimally violating the NCC (as it happens in \cite{Visser2019bis}, where $\rho + p_{\parallel} = k$, with $k$ constant and $k<0$ near the regular center) or even considering theories beyond General Relativity. Note that, in this case, departure from spherical symmetry is mandatory because, as we have shown in previous sections, only de Sitter (or Minkowski) cores are allowed within this symmetry.
\\
\\
\subsubsection{The many advantages of Nariai and Bertotti-Robinson cores}

Up to this point we have focused our attention on regular black holes with two different slices: $S^3$ at the $r\approx 0$ core and $\mathbb{R} \times S^2$ outside it. As we have shown along the manuscript, topology change between slices is not free of problems but it bring us some undesirable properties such as lack of global hyperbolicity and the existence of closed timelike curves. Although the first one could be solved using Geroch's prescription, causality violations seem to be unavoidable within most models of regular black holes.

Fortunately, the cosmological constant comes to our rescue. Let us elaborate this idea.

As shown by Bousso and Hawking \cite{Hawking1996,Bousso1}, the $S^1\times S^2$ topology of the spacelike sections of Reissner-Nordstr\"om-de Sitter spaces becomes evident when an appropriate coordinate change is performed (see the 
Appendix of \cite{Hawking1996} for the neutral case and \cite{Bousso1} for the charged case). In general, the radius of the $S^2$ sphere varies along the $S^1$ (the minimal
two-sphere corresponds to the black hole horizon and the maximal one to the cosmological horizon). Interestingly, for the charged Nariai case, the two-sphere radius is independent of the $S^1$-coordinate. In this sense, the spacelike slices of a charged Nariai solution can be thought as a as a ``perfect doughnut", while for generic Reissner-Nordstr\"om de Sitter solution it would be a ``wobbly doughnut". Therefore, if black holes immersed in a de Sitter space are considered, the topology of the spacelike slices are $S^1\times S^2$. This feature, which we remind the reader is a direct consequence of a non zero cosmological constant, avoids, upon regularizing these kind of black holes with a Nariai core near their center, topology changes between spacelike slices. Then, globally hyperbolic regular black holes satisfying Borde's theorem are not forbidden and they can be constructed in principle.
\\
\\
\indent 
Interestingly, as mentioned in the introduction, a very recent example of a classical mechanism giving place to a regular black hole with a Nariai center but violating the NEC has given in \cite{Mariam} by the use of three-form fields. 
A different example was presented in \cite{Melgarejo2020}, also very recently, based on a slight modification of the well-known spherically symmetric Hayward black hole \cite{Hayward} satisfying the NEC everywhere. Specifically, the proposed line element, which reads
 \begin{equation}
 \label{Hay}
ds^{2}=-\left(1-\frac{2 m r^2}{r^3+ 2 l^2 m}\right)dt^2+\frac{dr^2}{1-\frac{2 m r^2}{r^3+ 2 l^2 m}}+l^2 d\Omega^2,
 \end{equation}
\\
\\
has a Nariai core with slices $S^1\times S^2$, in complete agreement with Borde's theorem. The topology of the spatial slices at spacelike infinity is $\mathbb{R}\times S^2$ for the previous solution and, therefore, topology change appears. In order to bypass this issue, let us embed this geometry into a de Sitter space. The result is

 \begin{equation}
 \label{Hay2}
ds^{2}=-\left(1-\frac{2 m r^2}{r^3+ 2 l^2 m}-\frac{\Lambda}{3}r^2\right)dt^2+\frac{dr^2}{1-\frac{2 m r^2}{r^3+ 2 l^2 m}-\frac{\Lambda}{3}r^2}+\frac{1}{\Lambda} d\Omega^2.
 \end{equation}

This geometry is regular everywhere, it has a Nariai and a Schwarzschild de Sitter geometry for $r\approx 0$ and $r\approx \infty$, respectively, and satisfies the NEC everywhere. Therefore, it provides an interesting example of a geometry satisfying Borde's theorem without topology change.

If, on he contrary, electrovacuum black holes are considered, the topology of the spacelike slices are $\mathbb{R}\times S^2$ (we are assuming spherical horizons for simplicity). Therefore, if a Bertotti-Robinson core is assumed for the regularized black hole, topology change is avoided and global hyperbolicity can be restored in principle. However, Borde and Borde and Vilenkin's theorems imply, roughly speaking, that the NCC is violated if causal simplicity and the existence of trapped surfaces or reconverging light cones is guaranteed.
\\
\\
\indent 
As a summary of the present discussion, and based on the previous example, we can conclude with the following
\\
\\
\indent 
{\it Proposition}. Regular black holes satisfying Borde's theorem without topology changes are not forbidden.
\\
\\
\section{Final comments and conclusions}
\label{6}
Here we will briefly summarize the main novelties of our work together with some final comments and future work.
We have studied regular black holes from a global perspective looking for evading some of the well-known singularity theorems by using 
new ``reversed" results following the idea behind Borde's theorem. This strategy has allowed us to study the interplay between regularity, topology change and causality. It has been shown that, in general, the cores of
spherically symmetric and (locally) static regular black holes are $S^3$, $H^3$, $\mathbb{R}\times S^2$ or $S^1\times S^2$, depending whether a de Sitter, anti de Sitter, Bertotti-Robinson or Nariai geometries are employed to describe the slices at the regular center, showing the existence of more possibilities other than the well-known studied de Sitter and hollow cores. Some techniques from circle bundles have been employed to describe the transition between the core of the regular black hole and the rest of the slices in most of the cases considered in the literature. After studying the consequences of Geroch, Tipler and Borde and Vilenkin's theorems for topology change, we have shown that Nariai cores can be safely used to construct regular black holes satisfying Borde's theorem but, importantly, without topology changes. 
\\
\\
We end this work by pointing out some physical properties of the Nariai core, together with their differences with respect to to the usual de Sitter case, that are relevant for the singularity resolution \cite{Dadhich2001}. First, note that the gravitational charge density, $\rho + 3p$, is negative for de Sitter space and, therefore, it favours expansion. In this sense, the appearance of an effective cosmological constant gives place to an inflationary core instead of a singularity. Even more, de Sitter inflation is homogeneous, isotropic and it has no shear. On the contrary, Nariai cores do produce homogeneous but anisotropic and non-zero shear inflation. Intuitively, when shear is non-zero, the core is necessarily anisotropic and consequently it cannot be conformally flat. As a consequence, any spherically symmetric Nariai core is of Petrov type-D which, interestingly, coincides with the Petrov type of the asymptotic region of most of the regular solutions considered in the literature near spatial infinity. We consider that the intriguing relations between regularity, topology change and changes in the Petrov type, first suggested in \cite{Melgarejo2020}, deserve
further study.
\\
\\
As a final comment we would like to remind the reader that the present work has focused on regular solutions that are continuous
throughout the whole spacetime. Therefore, it remains to be seen what happens when regular solutions having boundary surfaces or surface layers joining the core and the enveloping metrics are considered.

\section{Acknowledgements}
P. B. acknowledges Ernesto Contreras for a critical reading of the manuscript. The author is funded by the Beatriz Galindo contract BEAGAL 18/00207 (Spain). P. B. dedicates this work to Anaís, Lucía, Inés and Ana. Comments and suggestions by three anonymous referees are gratefully acknowledged.

\end{document}